\documentclass[showpacs,aps,prc]{revtex4}
\usepackage[T1]{fontenc}
\usepackage[latin1]{inputenc}
\usepackage{graphicx}
\usepackage{amssymb}
\usepackage{verbatim}
\usepackage{epsfig}

\newcommand{\ba}{\begin{eqnarray}}
\newcommand{\ea}{\end{eqnarray}}

\begin{document}
\pagestyle{plain}

\title{Flavor content of nucleon form factors 
\footnote{Invited talk at XXX Symposium on Nuclear Physics, 
Hacienda Cocoyoc, Morelos, Mexico, January 3-6, 2007}}  
\author{R. Bijker}
\email{bijker@nucleares.unam.mx}
\affiliation{Departamento de Estructura de la Materia, 
Instituto de Ciencias Nucleares, 
Universidad Nacional Aut\'onoma de M\'exico, 
A.P. 70-543, 04510 M\'exico, D.F., M\'exico}

\author{E. Santopinto}
\email{santopinto@ge.infn.it}
\affiliation{I.N.F.N. and Dipartimento di Fisica,\\ via Dodecaneso 33,
Genova, I-16146, Italy}

\begin{abstract}
The flavor content of nucleon form factors is analyzed using two 
different theoretical approaches. The first is based on a 
phenomenological two-component model in which the external photon 
couples to both an intrinsic three-quark structure and a meson 
cloud via vector-meson dominance. The flavor content of the nucleon 
form factors is extracted without introducing any additional parameter. 
A comparison with recent data from parity-violating electron scattering 
experiments shows a good overall agreement for the strange form factors.  

A more microscopic approach is that of an unquenched quark model  
proposed by Geiger and Isgur which is based on valence quark plus 
glue dominance to which quark-antiquark pairs are added in 
perturbation. In the original version the importance of $s \bar{s}$ 
loops in the proton was studied. Here we present the formalism for a new 
generation of unquenched quark models which, among other extensions, 
includes the contributions of $u \bar{u}$ and $d \bar{d}$ loops. 
Finally, we discuss some preliminary results in the closure limit.   

\


\

Se analiza el contenido de extra\~neza de los factores de forma del nucle\'on 
usando dos m\'etodos te\'oricos distintos. El primero est\'a basado en un modelo 
fenomenol\'ogico de dos componentes en que el fot\'on se acopla tanto a una 
estructura intr{\'{\i}}nseca de tres cuarks como a una nube mes\'onica 
a trav\'es de la dominancia de mesones vectoriales. Se determina el contenido 
de sabor de los factores de forma del nucle\'on sin la necesidad de introducir 
alg\'un par\'ametro adicional. Una comparaci\'on con datos recientes de 
experimentos de dispersi\'on de electrones con violaci\'on de paridad muestra 
un buen ajuste para los factores de forma con extra\~neza.  

Un m\'etodo m\'as microsc\'opico es el de un modelo de cuarks 'unquenched'  
propuesto por Geiger e Isgur con base en la dominancia de cuarks de valencia 
m\'as gluones al cual se agregan pares de cuark-anticuark en perturbaci\'on. 
En la versi\'on original se estudi\'o la importancia de los lazos $s \bar{s}$ 
para el prot\'on. En este trabajo se presenta el formalismo de una nueva 
generaci\'on de modelos de cuarks 'unquenched' que, entre otras extensiones, 
incluye las contribuciones de los lazos $u \bar{u}$ y $d \bar{d}$. 
Finalmente, se discuten algunos resultados preliminares en el l{\'{\i}}mite 
de cerradura. 

\
 
\noindent
Descriptores: Bariones, factores de forma extra\~nos, dominancia de mesones vectoriales, 
modelos de cuarks.
\end{abstract}


\maketitle                   

\section{Introduction}

Recent experimental developments have made it possible to determine 
the flavor content of nucleon form factors. In particular the strange 
form factors of the proton can be obtained by combining asymmetry 
measurements in parity-violating electron scattering (PVES) 
\cite{Spayde,Maas,Happex,Armstrong} with either the electromagnetic form 
factors of the nucleon or with neutrino-proton scattering data \cite{Pate}.  
The first results from the SAMPLE, PVA4, HAPPEX and G0 collaborations 
have shown evidence for a nonvanishing strange quark contribution, 
albeit small, to the charge and magnetization distributions of the 
proton \cite{Young}. 

In the constituent quark model (CQM), the proton is described in terms 
of a $uud$ three-quark configuration. Therefore, nonvanishing strange 
form factors provide direct evidence for the presence of higher Fock 
components in the proton wave function (such as $uud-s \bar{s}$ 
configurations). The contribution of strange quarks to the nucleon is of 
special interest because it is exclusively part of the quark-antiquark 
sea. There is a wide variety of CQMs: {\it e.g.} the Isgur-Karl model 
\cite{IK}, the Capstick-Isgur model \cite{capstick}, the algebraic 
$U(7)$ model \cite{bil1,bil2}, the hypercentral model \cite{pl}, the chiral 
boson exchange model \cite{olof} and the Bonn instanton model \cite{bn}. 
Any of these models is able to reproduce the mass spectrum of baryon 
resonances reasonably well, but all of them show very similar deviations 
for other properties, such as for example the electromagnetic and strong 
decay widths of $\Delta(1232)$ and $N(1440)$, the spin-orbit splitting 
of $\Lambda(1405)$ and $\Lambda(1520)$, the transition form factors of 
$\Delta(1232)$, $N(1440)$, $N(1520)$, $N(1535)$ and $N(1680)$, and the 
large $\eta$ decay widths of the $N(1535)$, $\Lambda(1670)$ and 
$\Sigma(1750)$ resonances which are very close to the threshold for $\eta$ decay.  
In \cite{bil2} it was found that the main discrepancies occur for the 
low-lying $S$-wave states, specifically $N(1535)$, $\Lambda(1405)$, 
$\Lambda(1670)$, $\Lambda(1800)$ and $\Sigma(1750)$, which 
have masses close to the threshold of a meson-baryon decay channel. 
All of these results point towards the need to include exotic degrees 
of freedom ({\em i.e.} other than $qqq$), such as multiquark $q^4 \bar{q}$ 
or gluonic $q^3 g$ configurations. 
Another piece of evidence for quark-antiquark components in 
the proton comes from measurements of the 
$\bar{d}/\bar{u}$ asymmetry in the nucleon sea \cite{sea}. 
   
The role of higher Fock components in the CQM 
has been studied theoretically in a series of papers by Riska {\em et al.} 
\cite{riska} in which it was shown that an appropriate admixture 
of some $q^4 \bar{q}$ configurations may reduce the observed 
discrepancies between experiment and theory for several low-lying 
baryon resonances. In another CQM based approach by Isgur and 
collaborators, the effects of quark-antiquark pairs were included 
in a flux-tube breaking model based on valence-quark 
plus glue dominance to which $q \bar{q}$ pairs are added in 
perturbation \cite{mesons,baryons}. It was found necessary 
to sum over a large set of intermediate states in order to 
preserve the successes of the CQM, such as for example the OZI 
hierarchy \cite{OZI}. In \cite{deMelo}, a possible change of 
sign in the proton form factor ratio $\mu_p G_E^p/G_M^p$ is 
attributed to the interplay between the contribution from the 
elastic quark-photon vertex and the one from the pair 
production process, {\em i.e.} higher Fock components.  

The aim of the present contribution is to study the importance of 
quark-antiquark pairs in baryon spectroscopy. We start by analyzing 
the available experimental data on strange form factors in a  
phenomenological approach \cite{RB1} based on a two-component model 
of nucleon form factors in terms of an intrinsic structure ($qqq$ 
configuration) surrounded by a meson cloud ($q \bar{q}$ pairs) 
\cite{BI} from which the flavor content is extracted according 
to a procedure first introduced by Jaffe \cite{Jaffe}. 
Next we present an unquenched quark model which is a generalization 
of the flux-tube breaking model proposed by Geiger and Isgur \cite{baryons}. 
As a first application, we discuss the flavor decomposition of the 
spin of the ground state octet and decuplet baryons in the closure limit. 

\section{Two-component model} 

First we study the strange form factors of the nucleon in a phenomenological 
two-component model \cite{IJL,BI} in which it is assumed that the external photon 
couples both to an intrinsic three-quark structure described by the dipole form 
factor $g(Q^2)=1/(1+\gamma Q^2)^2$, and to a meson cloud via vector-meson 
($\rho$, $\omega$ and $\phi$) dominance (VMD). In the original VMD calculation 
\cite{IJL}, the Dirac form factor was attributed to both the intrinsic structure and 
the meson cloud, and the Pauli form factor entirely to the meson cloud. 
In \cite{BI}, it was shown that the addition of an intrinsic 
part to the isovector Pauli form factor, as suggested by studies of 
relativistic constituent quark models in the light-front approach 
\cite{frank}, considerably improves the results for the neutron electric and magnetic 
form factors. 

\subsection{Strange form factors}

Electromagnetic and weak form factors contain the information about the 
distribution of electric charge and magnetization inside the nucleon.  
These form factors arise from matrix elements of the corresponding 
vector current operators
\ba
\left< N \left| V_{\mu} \right| N \right> = \bar{u}_N \left[ 
F_1(Q^2) \, \gamma_{\mu} + \frac{i}{2M_N} F_2(Q^2) \, \sigma_{\mu\nu} q^{\nu} 
\right] u_N ~.
\ea
Here $F_{1}$ and $F_2$ are the Dirac and Pauli form factors, 
which are functions of the squared momentum transfer $Q^2=-q^2$. 
The electric and magnetic form factors, $G_{E}$ and $G_{M}$, are 
obtained from $F_{1}$ and $F_{2}$ by the relations $G_E=F_1-\tau F_2$ 
and $G_M=F_1 + F_2$ with $\tau=Q^2/4 M_N^2$. 

Since the intrinsic part is associated with the valence quarks of the 
nucleon, the strange quark content of the nucleon form factors arises 
from the (isoscalar) meson wave functions 
\ba
\left| \omega \right> &=& \cos \epsilon \left| \omega_0 \right> 
- \sin \epsilon \left| \phi_0 \right> ,
\nonumber\\
\left| \phi \right> &=& \sin \epsilon \left| \omega_0 \right> 
+ \cos \epsilon \left| \phi_0 \right> ,
\ea
where $\left| \omega_0 \right>=\left( u \bar{u} + d \bar{d} \right)/\sqrt{2}$ 
and $\left| \phi_0 \right> = s \bar{s}$ are the ideally mixed states. 
Under the assumption that the strange form factors have the same form as 
the isoscalar ones, the strange Dirac and Pauli form factors are expressed as the 
product of an intrinsic part $g(Q^2)$ and a contribution from the meson cloud as 
\cite{RB1}
\ba
F_{1}^{s}(Q^{2}) &=& \frac{1}{2}g(Q^{2})\left[ 
\beta_{\omega}^s \frac{m_{\omega}^{2}}{m_{\omega }^{2}+Q^{2}} 
+\beta_{\phi}^s \frac{m_{\phi}^{2}}{m_{\phi }^{2}+Q^{2}}\right] ~, 
\nonumber\\
F_{2}^{s}(Q^{2}) &=& \frac{1}{2}g(Q^{2})\left[ 
\alpha_{\omega}^s \frac{m_{\omega}^{2}}{m_{\omega }^{2}+Q^{2}}
+\alpha_{\phi}^s \frac{m_{\phi}^{2}}{m_{\phi }^{2}+Q^{2}}\right] ~,
\label{sff}
\ea
where the $\beta$'s and $\alpha$'s are related to the two independent isoscalar 
couplings $\beta_{\omega}$ and $\alpha_{\phi}$ \cite{RB1,RB2}
\ba
\beta_{\omega}^s &=& -\beta_{\phi}^s = 
-\sqrt{6} \, \frac{\sin \epsilon}{\sin(\theta_0+\epsilon)} \beta_{\omega} ~,
\nonumber\\
\alpha_{\omega}^s &=& -\sqrt{6} \, \frac{\sin \epsilon}{\sin(\theta_0+\epsilon)} 
\left( \mu_p + \mu_n -1 - \alpha_{\phi} \right) ~, 
\nonumber\\
\alpha_{\phi}^s &=& 
-\sqrt{6} \, \frac{\cos \epsilon}{\cos(\theta_0+\epsilon)} \alpha_{\phi} ~.  
\label{coef2}
\ea
Here $\tan \theta_0 = 1/\sqrt{2}$. The Dirac form factor $F_1^s$ is small due 
to canceling contributions of the $\omega$ and $\phi$ couplings which arise 
as a consequence of the fact that the strange (anti)quarks do not contribute 
to the electric charge $G_E^s(0)=F_1^s(0)=\beta_{\omega}^s+\beta_{\phi}^s=0$.  

The mixing angle $\epsilon$ can be determined from the decay properties of the 
$\omega$ and $\phi$ mesons as $\epsilon=0.053$ rad \cite{Jain}. 
Since the mixing angle is very small, it is worth examining the results for 
the strange Dirac and Pauli form factors in the absence of mixing. In the limit of 
zero mixing,  the strangeness contribution arises entire from the Pauli coupling of the 
$\phi$ meson $\alpha^s_{\phi}$
\ba
F_{1}^{s}(Q^{2}) &\rightarrow& 0 ~,
\nonumber\\
F_{2}^{s}(Q^{2}) &\rightarrow& \frac{1}{2}g(Q^{2}) \, 
\alpha_{\phi}^s \frac{m_{\phi}^{2}}{m_{\phi }^{2}+Q^{2}} ~. 
\label{ideal}
\ea

\subsection{Results}

In the present calculation of the strange form factors, 
the coefficient $\gamma$ in the intrinsic form factor and the isoscalar 
couplings, $\beta_{\omega}$ and $\alpha_{\phi}$, are taken from a study of the 
electromagnetic form factors of the nucleon. According to 
Eq.~(\ref{coef2}), the strange couplings can be determined from the 
isoscalar couplings: $\beta_{\phi}^s=-\beta_{\omega}^s=0.202$,  
$\alpha_{\phi}^s=0.648$ and $\alpha_{\omega}^s=-0.018$ \cite{RB1}.   

\begin{figure}[htb]
\centerline{\epsfig{file=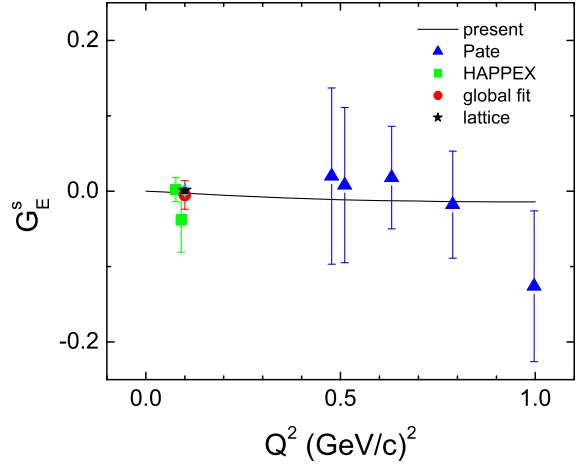,width=0.5\textwidth}} 
\caption[]{\small 
Comparison between theoretical and experimental values of the strange 
electric form factor. The experimental values are taken from \cite{Happex} 
(squares), a global fit \cite{Happex} (circle) and \cite{Pate} (triangles). 
The result from lattice QCD is indicated by a star \cite{leinweber2}.}
\label{GEs}
\end{figure}

\begin{figure}[htb]
\centerline{\epsfig{file=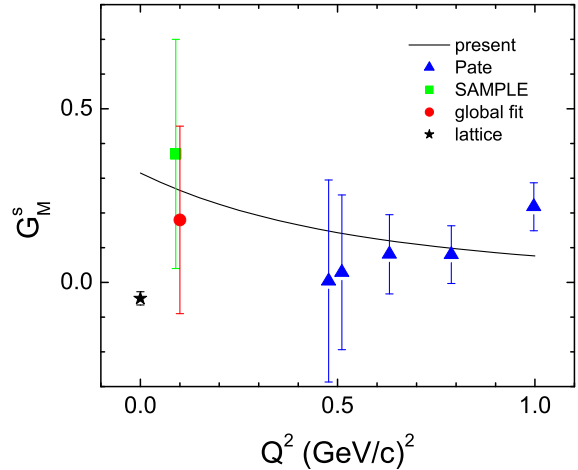,width=0.5\textwidth}} 
\caption[]{\small 
Comparison between theoretical and experimental values of the strange 
magnetic form factor. The experimental values are taken from \cite{Spayde} 
(square), a global fit \cite{Happex} (circle) and \cite{Pate} (triangles). 
The result from lattice QCD is indicated by a star \cite{leinweber1}.}
\label{GMs}
\end{figure}

Figs.~\ref{GEs} and \ref{GMs} show the strange electric and magnetic form 
factors as a function of $Q^2$. The qualitative features can be understood 
in the limit of ideally mixed mesons, {\it i.e.} zero mixing angle 
(in comparison with the value of $\epsilon=3.0^{\circ}$ 
used in Figs.~\ref{GEs} and \ref{GMs}). In this case, according to 
Eq.~(\ref{ideal}) the Dirac form factor vanishes. Hence the strange electric 
and magnetic form factors reduce to 
\ba
G_{E}^{s} &\rightarrow& - \tau F_{2}^{s} ~,
\nonumber\\  
G_{M}^{s} &\rightarrow& F_{2}^{s} ~.
\ea
The strange electric form factor 
is small since, for the range of $Q^2$ in Fig.~\ref{GEs}, the contribution 
from the Pauli form factor is suppressed by the factor $\tau=Q^2/4M_N^2$. 
 
The theoretical values are in good agreement with the recent experimental 
results of the HAPPEX Collaboration in which $G_E^s$ was determined in PVES 
from a $^{4}$He target \cite{Happex}, 
as well as with the result from an analysis of the world data 
$G_E^s(Q^2=0.1) = -0.005 \pm 0.019$ \cite{Happex} and from a lattice QCD calculation 
$G_E^s(0.1) = -0.009 \pm 0.005 \pm 0.003 \pm 0.027$ \cite{leinweber2}. 

The strange magnetic form factor $G_M^s=F_1^s + F_2^s$ is positive, since 
it is dominated by the contribution from the Pauli form factor. 
The SAMPLE experiment measured the parity-violating asymmetry at backward 
angles, which allowed  the strange magnetic form factor at 
$Q^2=0.1$ (GeV/c)$^2$ to be determined as $G_M^s=0.37 \pm 0.20 \pm 0.26 \pm 0.07$.  
An analysis of the world data at the same value of $Q^2$ gives 
$G_M^s = 0.18 \pm 0.27$ \cite{Happex}. 
The other experimental values of $G_E^s$ and $G_M^s$ in Figs.~\ref{GEs} 
and \ref{GMs} were obtained \cite{Pate} by combining the 
(anti)neutrino data from E734 \cite{Ahrens} with the parity-violating 
asymmetries from HAPPEX \cite{Happex} and G0 \cite{Armstrong}. 
The theoretical values are in good overall agreement with the 
experimental ones for the entire range $0 < Q^2 < 1.0$ (GeV/c)$^2$. 

The strange magnetic moment is calculated to be positive 
\ba
\mu_s = G_M^s(0) = \frac{1}{2} (\alpha_{\omega}^s+\alpha_{\phi}^{s}) 
= 0.315 \, \mu_N ~,
\ea
in units of the nuclear magneton, $\mu_N=e \hbar/2M_N c$. This value is in 
agreement with that of a recent analysis of the world data on strange form 
factors in the range of $0 < Q^2 < 3$ (GeV/c)$^2$ which gives 
$\mu_s = 0.37 \pm 0.79$ $\mu_N$ \cite{Young}. We note that the latter 
evaluation did not include the new HAPPEX data from \cite{Happex}.     
Theoretical calculations of the strange magnetic moment show a large variation, 
although most QCD-inspired models seem to favor a negative value in the range 
$-0.6 \lesssim \mu_s \lesssim 0.0$ $\mu_N$ \cite{beck}. 
Recent lattice-QCD calculations give small values, {\it e.g.}   
$0.05 \pm 0.06$ \cite{lewis} and $-0.046 \pm 0.019$ \cite{leinweber1}. 

Most experimental data on strange form factors correspond to a linear combination 
of electric and magnetic form factors $G_E^s + \eta G_M^s$. In \cite{RB1,RB2} 
it was shown that the calculated values in the two-component model are in good 
overall agreement with the experimental data from the PVA4 \cite{Maas}, HAPPEX 
\cite{Happex} and G0 collaborations \cite{Armstrong}. 
Finally, a flavor decomposition of the nucleon electromagnetic form factors 
shows that the contribution of the strange quarks to the proton form factors 
is small, being of the order of a few percent of the total \cite{RB2}. 

Another recent extension of the two-component model of the nucleon has been to 
the transition form factors of baryon resonances \cite{Wan}. 
The two-component model has the disadvantage that its applicability depends on 
the availability of a good and reliable set of experimental data to be able to 
determine the coefficients in the Dirac and Pauli form factors. The couplings 
between the intrinsic structure and the quark-antiquark pairs are obtained in 
a phenomenological rather than a dynamical way, in which these couplings would  
be the result of a specific interaction term. In the next section, we discuss 
the flux-tube breaking model, in which the effects of the higher Fock components 
are included via a $^{3}P_0$ coupling mechanism. 

\section{Flux-tube breaking model}

In the flux-tube model for hadrons, the quark potential model arises from an 
adiabatic approximation to the gluonic degrees of freedom embodied in the flux 
tube \cite{flux}. The impact of quark-antiquark pairs in meson spectroscopy has 
been studied in an elementary flux-tube breaking model \cite{mesons} in which 
the $q \bar{q}$ pair is created with the $^{3}P_0$ quantum numbers of the vacuum.  
Subsequently, it was shown by Geiger and Isgur \cite{OZI} that a "miraculous" 
set of cancellations between apparently uncorrelated sets of intermediate states 
occurs in such a way that they compensate each other and do not destroy the good 
CQM results for the mesons. In particular, the OZI hierarchy is preserved and 
there is a near immunity of the long-range confining potential, since the change 
in the linear potential due to the creation of quark-antiquark pairs in the string 
can be reabsorbed into a new strength of the linear potential, {\em i.e.} in a new 
string tension. As a result, the net effect of the mass shifts from pair creation 
is smaller than the naive expectation of the order of the strong decay widths. 
However, it is necessary to sum over  large towers of intermediate 
states to see that the spectrum of the mesons, after unquenching and renormalizing, 
is only weakly perturbed. An important conclusion is that no simple truncation of 
the set of meson loops is able to reproduce such results \cite{OZI}.

\begin{figure}
\centering
\setlength{\unitlength}{0.7pt}
\begin{tabular}{ccc}
\begin{picture}(120,160)(10,10)
\thicklines
\put( 40, 20) {\circle{5}}
\put( 60, 20) {\circle{5}}
\put( 80, 20) {\circle{5}}
\put( 20,120) {\circle{5}}
\put( 40,120) {\circle{5}}
\put( 60,120) {\circle{5}}
\put( 80,120) {\circle*{5}}
\put(100,120) {\circle{5}}
\put( 40, 22.5) {\line( 0,1){57.5}}
\put( 40, 80) {\line(-1,2){18.9}}
\put( 60, 22.5) {\line( 0,1){57.5}}
\put( 60, 80) {\line(-1,2){18.9}}
\put( 80, 22.5) {\line( 0,1){57.5}}
\put( 80, 80) {\line( 1,2){18.9}}
\put( 70,100) {\line(-1,2){ 8.9}}
\put( 70,100) {\line( 1,2){ 8.9}}
\put( 35,  5) {$q_1$}
\put( 55,  5) {$q_2$}
\put( 75,  5) {$q_3$}
\put( 15,130) {$q_1$}
\put( 35,130) {$q_2$}
\put( 55,130) {$q$}
\put( 75,130) {$\bar{q}$}
\put( 95,130) {$q_3$}
\end{picture}
& 
\begin{picture}(120,160)(10,10)
\thicklines
\put( 40, 20) {\circle{5}}
\put( 60, 20) {\circle{5}}
\put( 80, 20) {\circle{5}}
\put( 20,120) {\circle{5}}
\put( 40,120) {\circle{5}}
\put( 60,120) {\circle{5}}
\put( 80,120) {\circle*{5}}
\put(100,120) {\circle{5}}
\put( 40, 22.5) {\line( 0,1){37.5}}
\put( 40, 60) {\line( 1,1){20}}
\put( 60, 80) {\line(-1,2){18.9}}
\put( 60, 22.5) {\line( 0,1){37.5}}
\put( 60, 60) {\line(-1,1){20}}
\put( 40, 80) {\line(-1,2){18.9}}
\put( 80, 22.5) {\line( 0,1){57.5}}
\put( 80, 80) {\line( 1,2){18.9}}
\put( 70,100) {\line(-1,2){8.9}}
\put( 70,100) {\line( 1,2){8.9}}
\put( 35,  5) {$q_1$}
\put( 55,  5) {$q_2$}
\put( 75,  5) {$q_3$}
\put( 15,130) {$q_2$}
\put( 35,130) {$q_1$}
\put( 55,130) {$q$}
\put( 75,130) {$\bar{q}$}
\put( 95,130) {$q_3$}
\end{picture}
&
\begin{picture}(120,160)(10,10)
\thicklines
\put( 40, 20) {\circle{5}}
\put( 60, 20) {\circle{5}}
\put( 80, 20) {\circle{5}}
\put( 20,120) {\circle{5}}
\put( 40,120) {\circle{5}}
\put( 60,120) {\circle{5}}
\put( 80,120) {\circle*{5}}
\put(100,120) {\circle{5}}
\put( 40, 22.5) {\line( 0,1){37.5}}
\put( 40, 60) {\line( 2,1){40}}
\put( 80, 80) {\line( 1,2){18.9}}
\put( 60, 22.5) {\line( 0,1){37.5}}
\put( 60, 60) {\line(-1,1){20}}
\put( 40, 80) {\line(-1,2){18.9}}
\put( 80, 22.5) {\line( 0,1){37.5}}
\put( 80, 60) {\line(-1,1){20}}
\put( 60, 80) {\line(-1,2){18.9}}
\put( 70,100) {\line(-1,2){8.9}}
\put( 70,100) {\line( 1,2){8.9}}
\put( 35,  5) {$q_1$}
\put( 55,  5) {$q_2$}
\put( 75,  5) {$q_3$}
\put( 15,130) {$q_2$}
\put( 35,130) {$q_3$}
\put( 55,130) {$q$}
\put( 75,130) {$\bar{q}$}
\put( 95,130) {$q_1$}
\end{picture}
\\
\begin{picture}(120,160)(10,10)
\thicklines
\put( 40, 20) {\circle{5}}
\put( 60, 20) {\circle{5}}
\put( 80, 20) {\circle{5}}
\put( 20,120) {\circle{5}}
\put( 40,120) {\circle{5}}
\put( 60,120) {\circle{5}}
\put( 80,120) {\circle*{5}}
\put(100,120) {\circle{5}}
\put( 40, 22.5) {\line( 0,1){37.5}}
\put( 40, 60) {\line( 2,1){40}}
\put( 80, 80) {\line( 1,2){18.9}}
\put( 60, 22.5) {\line( 0,1){57.5}}
\put( 60, 80) {\line(-1,2){18.9}}
\put( 80, 22.5) {\line( 0,1){37.5}}
\put( 80, 60) {\line(-2,1){40}}
\put( 40, 80) {\line(-1,2){18.9}}
\put( 70,100) {\line(-1,2){8.9}}
\put( 70,100) {\line( 1,2){8.9}}
\put( 35,  5) {$q_1$}
\put( 55,  5) {$q_2$}
\put( 75,  5) {$q_3$}
\put( 15,130) {$q_3$}
\put( 35,130) {$q_2$}
\put( 55,130) {$q$}
\put( 75,130) {$\bar{q}$}
\put( 95,130) {$q_1$}
\end{picture}
& 
\begin{picture}(120,160)(10,10)
\thicklines
\put( 40, 20) {\circle{5}}
\put( 60, 20) {\circle{5}}
\put( 80, 20) {\circle{5}}
\put( 20,120) {\circle{5}}
\put( 40,120) {\circle{5}}
\put( 60,120) {\circle{5}}
\put( 80,120) {\circle*{5}}
\put(100,120) {\circle{5}}
\put( 40, 22.5) {\line( 0,1){37.5}}
\put( 40, 60) {\line( 1,1){20}}
\put( 60, 80) {\line(-1,2){18.9}}
\put( 60, 22.5) {\line( 0,1){37.5}}
\put( 60, 60) {\line( 1,1){20}}
\put( 80, 80) {\line( 1,2){18.9}}
\put( 80, 22.5) {\line( 0,1){37.5}}
\put( 80, 60) {\line(-2,1){40}}
\put( 40, 80) {\line(-1,2){18.9}}
\put( 70,100) {\line(-1,2){8.9}}
\put( 70,100) {\line( 1,2){8.9}}
\put( 35,  5) {$q_1$}
\put( 55,  5) {$q_2$}
\put( 75,  5) {$q_3$}
\put( 15,130) {$q_3$}
\put( 35,130) {$q_1$}
\put( 55,130) {$q$}
\put( 75,130) {$\bar{q}$}
\put( 95,130) {$q_2$}
\end{picture}
&
\begin{picture}(120,160)(10,10)
\thicklines
\put( 40, 20) {\circle{5}}
\put( 60, 20) {\circle{5}}
\put( 80, 20) {\circle{5}}
\put( 20,120) {\circle{5}}
\put( 40,120) {\circle{5}}
\put( 60,120) {\circle{5}}
\put( 80,120) {\circle*{5}}
\put(100,120) {\circle{5}}
\put( 40, 22.5) {\line( 0,1){57.5}}
\put( 40, 80) {\line(-1,2){18.9}}
\put( 60, 22.5) {\line( 0,1){37.5}}
\put( 60, 60) {\line( 1,1){20}}
\put( 80, 80) {\line( 1,2){18.9}}
\put( 80, 22.5) {\line( 0,1){37.5}}
\put( 80, 60) {\line(-1,1){20}}
\put( 60, 80) {\line(-1,2){18.9}}
\put( 70,100) {\line(-1,2){8.9}}
\put( 70,100) {\line( 1,2){8.9}}
\put( 35,  5) {$q_1$}
\put( 55,  5) {$q_2$}
\put( 75,  5) {$q_3$}
\put( 15,130) {$q_1$}
\put( 35,130) {$q_3$}
\put( 55,130) {$q$}
\put( 75,130) {$\bar{q}$}
\put( 95,130) {$q_2$}
\end{picture}
\end{tabular}
\vspace{15pt}
\caption[]{\small Quark line diagrams for $A \rightarrow B C$ with 
$q_1 q_2 q_3 = uud$ and $q \bar{q} = s \bar{s}$}
\label{diagrams}
\end{figure}
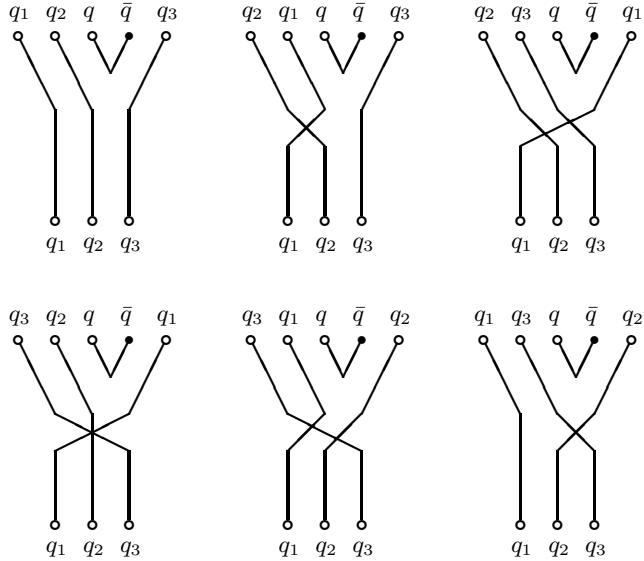

The extension of the flux-tube breaking model to baryons requires a proper treatment 
of the permutation symmetry between identical quarks. As a first step, Geiger and Isgur 
investigated the importance of $s \bar{s}$ loops in the proton by taking into account 
the contribution of the six different diagrams of Fig.~\ref{diagrams} with 
$q \bar{q}=s \bar{s}$ and $q_1 q_2 q_3 = uud$, and by using 
harmonic oscillator wave functions for the baryons and mesons \cite{baryons}. 
In the conclusions, the authors emphasized: {\em It also seems very worthwhile 
to extend this calculation to $u \bar{u}$ and $d \bar{d}$ loops. Such an extension 
could reveal the origin of the observed violations of the Gottfried sum rule and 
also complete our understanding of the origin of the spin crisis.} 
In this contribution, we take up this challenge and present the first results 
for some generalizations of the formalism of \cite{baryons} which make it now 
possible to study the quark-antiquark contributions 
\begin{itemize}
\item for any initial baryon resonance
\item for any flavor of the quark-antiquark pair
\item for any model of baryons and mesons, as long as their wave functions are 
expressed on the basis of the harmonic oscillator. 
\end{itemize}
The problem of the permutation symmetry between identical quarks has been solved by 
means of group-theoretical techniques. In this way, the quark-antiquark contribution 
can be calculated for any initial baryon $q_1 q_2 q_3$ (ground state or resonance) 
and for any flavor of the quark-antiquark pair $q \bar{q}$ (not only $s\bar{s}$, 
but also $u\bar{u}$ and $d\bar{d}$). 

\begin{figure}[ht]
\centering
\setlength{\unitlength}{0.7pt}
\begin{picture}(120,150)(50,0)
\thicklines
\put(100, 20) {\line(0,1){110}}
\put(110, 20) {\line(0,1){110}}
\put(120, 20) {\line(0,1){30}}
\put(120, 60) {\line(0,1){30}}
\put(120,100) {\line(0,1){30}}
\put(120, 75) {\oval(30,30)[br]}
\put(120, 75) {\oval(30,30)[tr]}
\put(120, 75) {\oval(50,50)[br]}
\put(120, 75) {\oval(50,50)[tr]}
\put( 60, 70) {B}
\put(170, 70) {C}
\end{picture}
\caption[]{\small One-loop diagram}
\label{loop}
\end{figure}
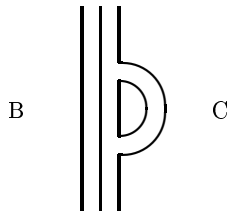

Here we adopt the unquenched quark model of \cite{baryons}, which is based on 
an adiabatic treatment of the flux-tube dynamics to which the $q \bar{q}$ pair 
creation with vacuum quantum numbers is added as a perturbation. 
The pair-creation mechanism is inserted at the quark level and the one-loop diagrams 
are calculated by summing over a complete set of intermediate states (see Fig.~\ref{loop}). 
Under these assumptions, the baryon wave function is, to leading order in pair 
creation, given by 
\begin{eqnarray} 
\mid \psi_A \rangle \;=\; {\cal N} \left[ \mid A \rangle 
+ \sum_{q BC l J} \int d \vec{k} \, \mid BC \vec{k} \, l J \rangle \, 
\frac{ \langle BC \vec{k} \, l J \mid h_{q \bar{q}}^{\dagger} \mid A \rangle } 
{M_A - E_B - E_C} \right] ~,
\end{eqnarray}
where $h_{q \bar{q}}^{\dagger}$ is the $^{3}P_0$ quark-antiquark pair-creation 
operator \cite{baryons}, $A$ is the initial baryon, $B$ and $C$ denote the intermediate 
baryon and meson, $\vec{k}$ and $l$ the relative radial momentum and orbital angular 
momentum of $B$ and $C$, and $J$ is the total angular momentum 
$\vec{J} = \vec{J}_{BC} + \vec{l} = \vec{J}_B + \vec{J}_C + \vec{l}$. 
The sum over $q$ denotes the sum over all flavors of the $q \bar{q}$ pair. 

In general, matrix elements of an observable $\hat{\cal O}$ can be expressed as 
\begin{eqnarray}
{\cal O} \;=\; \langle \psi_A \mid \hat{\cal O} \mid \psi_A \rangle 
\;=\; {\cal O}_{\rm valence} + {\cal O}_{\rm sea} ~,
\end{eqnarray}
where the first term denotes the contribution from the valence quarks  
\begin{eqnarray}
{\cal O}_{\rm valence} &=& {\cal N}^2 \langle A \mid \hat{\cal O} \mid A \rangle 
\end{eqnarray}
and the second term that from the quark-antiquark pairs
\begin{eqnarray}
{\cal O}_{\rm sea} &=& {\cal N}^2 \sum_{q BC l J} \int d \vec{k} \,
\sum_{q' B'C' l' J'} \int d \vec{k}^{\, \prime} \,
\frac{ \langle A \mid h_{q' \bar{q}'} \mid B' C' \vec{k}^{\, \prime} \, l' J' \rangle } 
{M_A - E_{B'} - E_{C'}} 
\nonumber\\
&& \hspace{2cm} \langle B' C' \vec{k}^{\, \prime} \, l' J' \mid \hat{\cal O}  
\mid B C \vec{k} \, l J \rangle \, 
\frac{ \langle B C \vec{k} \, l J \mid h_{q \bar{q}}^{\dagger} \mid A \rangle } 
{M_A - E_B - E_C} ~.
\label{me}
\end{eqnarray}
The sum in Eq.~(\ref{me}) is over a complete set of intermediate 
states, rather than just a few low-lying states. Not only does this have a significant 
impact on the numerical result, but it is necessary for consistency with the OZI-rule 
and the success of CQM's in hadron spectroscopy. We have developed an algorithm based 
upon group-theoretical techniques to generate the intermediate states with the correct 
permutational symmetry for any model of hadrons.  
Therefore, the sum over intermediate states can be perfomed up to saturation, and not 
just for the first few shells as in \cite{baryons}. 

\subsection{Closure limit}

The evaluation of the contribution of the quark-antiquark pairs simplifies considerably 
in the closure limit, which arises when the energy denominators do not depend strongly 
on the quantum numbers of the intermediate states in Eq.~(\ref{me}). In this case, the 
sum over the complete set of intermediate states can be solved by closure and the 
contribution of the quark-antiquark pairs to the matrix element reduces to 
\begin{eqnarray}
{\cal O}_{\rm sea} &\propto& \sum_{q q'} \langle A \mid h_{q' \bar{q}'} \, \hat{\cal O} \,   
h_{q \bar{q}}^{\dagger} \mid A \rangle ~.
\end{eqnarray}
Especially when combined with symmetries, the closure limit not only provides simple 
expressions for the relative flavor content of physical observables, but also can give 
further insight into the origin of cancellations between the contributions from different 
intermediate states. 

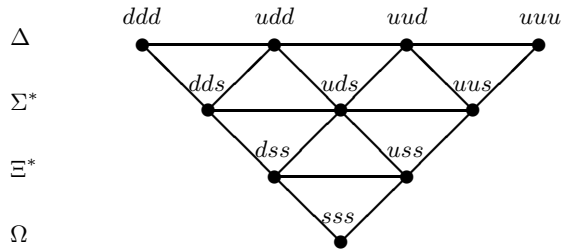
\begin{figure}[ht]
\centering
\setlength{\unitlength}{0.5pt}
\begin{picture}(440,200)(20,80)
\thicklines
\put(100,250) {\line(1,0){300}}
\put(150,200) {\line(1,0){200}}
\put(200,150) {\line(1,0){100}}

\put(150,200) {\line(1,1){ 50}}
\put(200,150) {\line(1,1){100}}
\put(250,100) {\line(1,1){150}}

\put(250,100) {\line(-1,1){150}}
\put(300,150) {\line(-1,1){100}}
\put(350,200) {\line(-1,1){ 50}}

\multiput(100,250)(100,0){4}{\circle*{10}}
\multiput(150,200)(100,0){3}{\circle*{10}}
\multiput(200,150)(100,0){2}{\circle*{10}}
\put(250,100){\circle*{10}}

\put( 85,265){$ddd$}
\put(185,265){$udd$}
\put(285,265){$uud$}
\put(385,265){$uuu$}
\put(135,215){$dds$}
\put(235,215){$uds$}
\put(335,215){$uus$}
\put(185,165){$dss$}
\put(285,165){$uss$}
\put(235,115){$sss$}

\put(0,250){$\Delta$}
\put(0,200){$\Sigma^{\ast}$}
\put(0,150){$\Xi^{\ast}$}
\put(0,100){$\Omega$}

\end{picture}
\caption[]{\small Ground state decuplet baryons}
\label{decuplet}
\end{figure}

As an example, we discuss some preliminary results for the operator 
\begin{eqnarray}
\Delta q \;=\; 2 \langle S_z(q) + S_z(\bar{q}) \rangle ~,
\end{eqnarray}
which determines the fraction of the baryon's spin carried by each one of the 
flavors $u$, $d$ and $s$. First, we consider the ground state decuplet baryons 
with $^{4}10 [56,0^+]_{3/2}$ of Fig.~\ref{decuplet}. 
Since the three-quark configuration of the $\Delta$ resonances does not 
contain strange quarks, the contribution of the $s \bar{s}$ pairs to the spin 
$\Delta s$ vanishes in the closure limit. The same holds for the contribution 
of $d \bar{d}$ pairs to the $\Delta^{++}$, $\Sigma^{\ast \, +}$, $\Xi^{\ast \, 0}$ 
and $\Omega^{-}$ resonances, and that of $u \bar{u}$ pairs to the $\Delta^{-}$, 
$\Sigma^{\ast \, -}$, $\Xi^{\ast \, -}$ and $\Omega^{-}$ resonances. Moreover, 
in the closure limit the relative contribution of the quark flavors from the 
quark-antiquark pairs to the baryon spin is the same as that from the valence quarks
\begin{eqnarray}
\Delta u_{\rm sea} : \Delta d_{\rm sea} : \Delta s_{\rm sea} \;=\; 
\Delta u_{\rm valence} : \Delta d_{\rm valence} : \Delta s_{\rm valence} ~.
\end{eqnarray}
This property is a consequence of the spin-flavor symmetry of the ground state 
baryons and holds for both the decuplet with quantum numbers $^{4}10 [56,0^+]_{3/2}$ 
and the octet with $^{2}10 [56,0^+]_{1/2}$. 
Table~\ref{baryonspin} shows the relative contributions of $\Delta u$, $\Delta d$ 
and $\Delta s$ to the spin of the ground state baryons in the closure limit. 
At a qualitative level, a vanishing closure limit explains the phenomenological 
success of CQMs. As an example, the strange content of the proton which vanishes 
in the closure limit is expected to be small, in agreement with the experimental 
data from PVES (for the most recent data see \cite{Happex,Armstrong}).  
Moreover, the results in Table~\ref{baryonspin} impose very stringent conditions 
on the numerical calculations, since each entry involves the sum over a complete set 
of intermediate states. Therefore, the closure limit provides a highly nontrivial 
test which involves both the spin-flavor sector, the permutation symmetry, the 
construction of a complete set of intermediate states and the 
implementation of the sum over all of these states. 

\begin{table}[ht]
\centering
\caption[]{\small Relative contributions of $\Delta u$, $\Delta d$ and $\Delta s$ 
in the closure limit to the spin of the ground state octet and decuplet baryons}   
\label{baryonspin}
\begin{tabular}{ccrcrcrcccccc}
\hline
& & & & & & & & & & & & \\
$qqq$ & $^{2}8[56,0^+]$  & $\Delta u$ &:& $\Delta d$ &:& $\Delta s$ 
      & $^{4}10[56,0^+]$ & $\Delta u$ &:& $\Delta d$ &:& $\Delta s$ \\
& & & & & & & & & & & & \\
\hline
& & & & & & & & & & & & \\
$uuu$ & & & & & & & $\Delta^{++}$ & 9 &:& 0 &:& 0 \\
$uud$ & $p$ & 4 &:& $-1$ &:& 0 
                  & $\Delta^{+}$ & 6 &:& 3 &:& 0 \\
$udd$ & $n$ & $-1$ &:& 4 &:& 0 
                  & $\Delta^{0}$ & 3 &:& 6 &:& 0 \\
$ddd$ & & & & & & & $\Delta^{-}$ & 0 &:& 9 &:& 0 \\
$uus$ & $\Sigma^+$ & 4 &:& 0 &:& $-1$ 
                  & $\Sigma^{\ast \, +}$ & 6 &:& 0 &:& 3 \\
$uds$ & $\Sigma^0$ & 2 &:& 2 &:& $-1$ 
                  & $\Sigma^{\ast \, 0}$ & 3 &:& 3 &:& 3 \\
      & $\Lambda$ & 0 &:& 0 &:& 3 && & & & & \\
$dds$ & $\Sigma^-$ & 0 &:& 4 &:& $-1$ 
                  & $\Sigma^{\ast \, -}$ & 0 &:& 6 &:& 3 \\
$uss$ & $\Xi^0$ & $-1$ &:& 0 &:& 4 
                  & $\Xi^{\ast \, 0}$ & 3 &:& 0 &:& 6 \\
$dss$ & $\Xi^-$ & 0 &:& $-1$ &:& 4 
                  & $\Xi^{\ast \, -}$ & 0 &:& 3 &:& 6 \\
$sss$ & & & & & & & $\Omega^{-}$ & 0 &:& 0 &:& 9 \\
& & & & & & & & & & & & \\
\hline
\end{tabular}
\end{table}

\section{Summary, conclusions and outlook}

In summary, we have discussed the importance of quark-antiquark pairs in baryon 
spectroscopy. A direct handle on these 
higher Fock components is provided by PVES experiments, which have shown evidence 
for a nonvanishing strange quark contribution, albeit small, to the charge and 
magnetization distributions of the proton. 

In the first part of this contribution, we reviewed an analysis of the recent 
data on strange form factors in a phenomenological two-component model which 
consists of an intrinsic (valence quark) structure surrounded by a meson cloud 
(quark-antiquark pairs) \cite{RB1}. It is important to note that the flavor content 
of the nucleon form factors is  extracted without introducing any additional 
parameter. All couplings were determined from a previous study of a simultaneous 
fit to the electromagnetic form factors of the nucleon (as a matter of fact, 
the static condition on the strange electric form factor $G_E^s(0)=0$ puts 
an additional constraint on the parameters, thus effectively reducing the number 
of independent parameters by one \cite{RB1}). A comparison with recent data 
from PVES experiments shows a good overall agreement 
for both the strange electric and magnetic form factors for the range of 
$0 < Q^2 < 1$ (GeV/c)$^2$. Therefore, one may conclude that the two-component model 
provides a simultaneous and consistent description of the electromagnetic 
and weak vector form factors of the nucleon. Future 
experiments on parity-violating electron scattering at backward angles 
(PVA4 and G0 \cite{g0back}) and neutrino scattering (FINeSSE \cite{finesse}) 
will make it possible to disentangle the contributions of the different 
quark flavors to the electric, magnetic and axial form factors, and thus 
to gain new insight into the complex internal structure of the nucleon. 

In the second part, we discussed the first results from a more microscopic approach 
to include the effects of the quark-antiquark pairs. The method is based on the 
dominance of valence quarks and gluon dynamics to which the $q \bar{q}$ pairs 
are added as a perturbation. The ensuing flux-tube breaking model was originally 
introduced by Kokoski and Isgur for mesons \cite{mesons} and later extended by 
Geiger and Isgur to $s \bar{s}$ loops in the proton \cite{baryons}. In this 
contribution, we presented a new generation of unquenched quark models for 
baryons by including, in addition to $s \bar{s}$ loops, the contributions of 
$u \bar{u}$ and $d \bar{d}$ loops as well. As an illustration, we applied the 
closure limit of the model - in which all intermediate states are assumed to be 
degenerate - to the flavor decomposition of the spin of the ground 
state octet and decuplet baryons. In this case, it was found that the relative 
contributions of the quark flavors from the $q \bar{q}$ pairs are the same as that 
from the valence quarks. 

The present formalism is, obviously within the assumptions of the approach, valid 
for any initial baryon, any flavor of the $q \bar{q}$ pairs and any model of hadron 
structure. In future work, it will be applied systematically to study several 
problems in light baryon spectroscopy, such as the spin crisis of the proton, 
the electromagnetic and strong couplings, the electromagnetic elastic 
and transition form factors of baryon resonances, their sea quark content and 
their flavor decomposition \cite{new}.  

\section*{Acknowledgments}
This work was supported in part by a grant from CONACYT, Mexico 
and in part by I.N.F.N., Italy.

\end{document}